\documentstyle[epsf]{l-aa}

\def\Etal#1{et al.\ (\cite{#1})}	%avec  parentheses
\def\etal#1{et al.\ \cite{#1}}
\def\ie{i.e.}
\def\fig#1{Fig.\,~\protect\ref{#1}}

\def\eq#1{Eq.\,~\protect\ref{#1}}

\def\sec#1{Sect.\,~\protect\ref{#1}}
\def\Sec#1{Section\,~\protect\ref{#1}}

\def\eg{e.g.}

\def\dh{$(^2{\rm H}/^1{\rm H})$}
\def\dhj{$(^2{\rm H}/^1{\rm H})_{\cal J} $}
\def\dhi{$(^2{\rm H}/^1{\rm H})_{\rm IMS}$}
\def\dhp{$(^2{\rm H}/^1{\rm H})_{\rm p}$}
\def\hethp{$(^3{\rm He}/^1{\rm H})_{\rm p}$}
\def\he{$(^3{\rm He}/^4{\rm He})$}
\def\hesw{$(^3{\rm He}/^4{\rm He})_{\rm SW}$}
\def\hep{$(^3{\rm He}/^4{\rm He})_{\rm p}$}
\def\hes{$(^3{\rm He}/^4{\rm He})_\odot$}
\def\hej{$(^3{\rm He}/^4{\rm He})_{\cal J} $}
\def\hehp{$(^4{\rm He}/^1{\rm H})_{\rm p}$}
\def\heh {$(^4{\rm He}/^1{\rm H})$}

\begin{document}

\thesaurus{\bf 02.14.1, 06.01.1, 07.19.1}

\title{A Reestimate of the Protosolar \dhp\ ratio from \hesw\ solar wind
measurements}

\author{D. Gautier\inst{1}\and P. Morel\inst{2}}

\offprints{gautier@meudon.obspm.fr}

\institute{DESPA, URA CNRS 264, Observatoire de Paris-Meudon, 92195 Meudon
Principal CEDEX, France.
\and D\'epartement Cassini, UMR CNRS 6529, Observatoire de la C\^ote
d'Azur, BP 4229, 06304 Nice CEDEX 4, France.}

\date{Received date; accepted date}

\maketitle

\markboth{D. Gautier \& P. Morel: A Reestimate of the Protosolar \dh\ ratio}
{D. Gautier \& P. Morel: A Reestimate of the Protosolar \dh\ ratio}

\begin{abstract}
 We reanalyze the inference of the protosolar abundance of deuterium
made by Geiss (\cite{g93}) from measurements of \hesw\ in the solar wind.
We use an evolutionary solar model with
microscopic diffusion, constrained to fit the present age, radius and
luminosity,
as well as the observed ratio of heavy elements to hydrogen. The protosolar
\dhp\ is obtained from the best fit of \hesw.
Taking for the protosolar \hep\ the value
measured in Jupiter by the Galileo probe (Niemann \etal{ni}), we derive
\dhp$=(3.01\pm 0.17)\times 10^{-5}$. Compared to the present
interstellar medium value
(Linsky \etal{li}), this result is compatible with models of
the chemical evolution of the Galaxy in the solar neighborhood; it is
also marginally compatible with the Jovian \dhj$=(5\pm 2)\times 10^{-5}$
ratio measured by Galileo.
\end{abstract}


\section{Introduction}\label{sec:in}
Since the pioneering work
of Geiss \& Reeves (\cite{ger}), a number of papers
have been devoted to the fundamental question of the estimate of
the protosolar abundance of deuterium as reviewed by Geiss
(\cite{g93}), hereafter referred as G93.
The current value of \dhp$=(2.6\pm 1.0)\times 10^{-5}$,
is that derived by G93
from a reanalysis of the measurements of \hesw\ in the Solar Wind (SW).

At the end of the pre-main sequence
the young Sun is still fully convective with a temperature, at center,
of a few million K; the mixing makes that all the deuterium
is converted to $^3{\rm He}$ via the
reaction $^2{\rm H}(p,\gamma)^3{\rm He}$. As pointed out by
Geiss \& Reeves (\cite{ger}), if \heh$\equiv0.1$ is no longer changed at the
surface then, \dhp\ is equal (G93) to \hehp\ times the difference between
\hes\ presently observed at the surface and the protosolar \hep:
\begin{displaymath}%\label{eq:gr}
\left(\frac{^2{\rm H}}{^1{\rm H}}\right)_{\rm p}=
\left(\frac{^4{\rm He}}{^1{\rm H}}\right)_{\rm p}\left[
\left(\frac{^3{\rm He}}{^4{\rm He}}\right)_\odot-
\left(\frac{^3{\rm He}}{^4{\rm He}}\right)_{\rm p}
\right].
\end{displaymath}
\hes\ is measured only in the solar wind, from the data available
G93 concluded to \hesw$=(4.5\pm 0.4)\times 10^{-4}$.
However G93 analyzing the various processes which may fractionate
the helium isotopes in the interior and in the atmosphere of the Sun,
has estimated that the actual ratio to be considered
for inferring \dhp\ is the solar wind value divided by a factor
$1.1\pm 0.2$, resulting in \hes$=(4.1\pm 1.0)\times 10^{-4}$.
\hep\ has been estimated by G93 from meteoritic
data to be equal to  \hep$=1.5\times 10^{-4}$.

An other way to determine the protosolar deuterium abundance is to evaluate
\dhj\ in
Jupiter. This approach is based on the fact that Jupiter is mainly made of
hydrogen which originates from the primordial solar nebula. According to
current models of formation of Jupiter, some amount of ices, presumably
enriched in deuterium with respect to the protosolar value, may have been
mixed
to hydrogen during the planetary formation, but their mass is currently
considered as too small
to have significantly increased \dhj\ (Hubbard \& MacFarlane \cite{hmf}).
Although the
various estimates of \dhj\ from remote sensing observations of HD
and of ${\rm CH}_3{\rm H}$ are uncertain, it was generally
considered, prior to the arrival of the Galileo mission to Jupiter,
that the Jovian deuterium abundance is consistent with the G93 value
(Lecluse \etal{le}).
The preliminary result of the mass spectrometer aboard the Galileo
atmospheric probe -- \dhj$=(5\pm 2)\times 10^{-5}$ (Niemann \etal{ni}) --
is however substantially higher than the G93 value. This result,
if confirmed, would have profound cosmological implications.

Compared to the present day
InterStellar Medium (ISM) value of \dhi$=(1.47-1.72)\times 10^{-5}$
(Linsky \etal{li}), it suggests a deuterium destruction 
in 4.55\,Gyr, the solar age, inconsistent with
current models of chemical evolution of galaxies. Alternatively, it might
be that \dhi\ varies upon the line of sight so that,
the concept of a well defined ISM deuterium abundance is not reliable,
as advocated by some authors \eg, Ferley \Etal{fe}.
A third possibility is that
the current protosolar value \dhp, is underestimated.

Two circumstances prompted us to reanalyze the determination of G93:
{\em -- First}, Niemann \Etal{ni} have
measured in Jupiter \hej$=1.1\times 10^{-4}$ which must be the
exact ratio in the primitive solar nebula. As pointed out by these authors,
this results in an increase of the estimation of \dhp.
{\em -- Second,} since 1993 the introduction of microscopic diffusion
in solar modeling
have significantly increased the accuracy of models (\eg, Basu \etal{bcd}).
A consequence of the microscopic diffusion is the change, with
respect to time, of surface values of \heh\ and \he.
A result directly relevant to this Letter is
the inferred helium mass fraction at the surface namely
$Y_\odot=0.242\pm 0.003$ (P\'erez Hern\'andez \& Christensen-Dalsgaard
\cite{hc})
substantially less than the protosolar value $Y_{\rm p}=0.275$
(Bahcall \& Pinsonneault \cite{bp}). Therefore,
we propose here to complete the approach of G93 by using a solar evolutionary
model with microscopic diffusion.
The modeling of the solar evolution and the data employed are
described in \Sec{sec:sm};
\Sec{sec:rd} is devoted to results and discussions. We conclude in \Sec{sec:c}.

\section{ Solar modeling and data}\label{sec:sm}
The calibrated solar models discussed in this paper include
the pre-main sequence evolution.
The calibration consists in the adjustment of,
i) the ratio of the mixing-length to the
pressure scale height, ii) the initial mass fraction $X_{\rm p}$
of hydrogen and,
iii) the initial protosolar mass fraction of heavy elements to hydrogen,
in order that the models have at present solar age, the observed
luminosity, radius (Guenter \etal{g})
and mass fraction of heavy elements to hydrogen
(Grevesse \& Noels \cite{gn}).
The models have been computed using the code CESAM
(Morel \cite{mp96}). The relevant features are:
the changes due to nuclear reactions, microscopic diffusion and convective
mixing are explicitly computed for
$^1{\rm H}$, $^2${\rm H}, $^3${\rm He}, $^4${\rm He}, $^7$Li, $^7$Be,
$^{12}$C, $^{13}$C,
$^{14}$N, $^{15}$N, $^{16}$O and $^{17}$O which enter into
the most important nuclear reactions of the PP+CNO cycles;
the protosolar abundance of each heavy element is derived
from $X_{\rm p}$
according to the nuclide abundances of Anders \& Grevesse (\cite{ag});
\hep\ is taken equal to \hej$=1.1\times 10^{-4}$ as measured by the
Galileo probe into Jupiter;
the microscopic diffusion coefficients of Michaud \& Proffitt (\cite{mipr})
are used, all species, but $^1{\rm H}$ and
$^4{\rm He}$, are trace elements.
\dhp\ is taken as a parameter and constrained to fit \hes.

As pointed out by G93, $^3$He could be somewhat favored over $^4$He in the
solar wind acceleration process (Burgi \& Geiss \cite{bg}). However,
the first results of the SWICS instrument aboard the Ulysses spacecraft
(Bodmer \etal{bo}) do not provide evidence that such a fractionation
could exceed the uncertainties on the measurements of the helium
isotopic ratio. Moreover, the chromospheric process is not expected
to result in a significant mass fractionation (G93).
Therefore we assume that \hes\ is equal,
within uncertainties, to \hesw.
\begin{figure}
\epsfxsize 8.cm
\epsfbox[25 18 587 524]{fig_2.ps}
\caption[]{
Profiles of \he\ in envelopes of
solar models computed with \dhp: $3.2\,10^{-5}$ (thick),
$3.6\,10^{-5}$ (medium) and  $2.8\,10^{-5}$ (thin).
The observed \hesw\ from APOLLO (1), ISEEE-3 (2) and Ulysses (3)
are plotted with their error bars.
}\label{fig:2}
\end{figure}
The values of \hesw\ obtained by three
experiments on three different space missions are:
\begin{description}
 \item{(1)}  Apollo \hesw$=(4.25\pm 0.21)\times 10^{-4}$ (Geiss \etal{ge}),
 \item{(2)} ISSEE-3 \hesw$=(4.88\pm 0.48)\times 10^{-4}$ (Coplan \etal{co},
Bochsler \cite{boc}),
 \item{(3)} Ulysses \hesw$=(4.4\pm 0.4)\times 10^{-4}$ (Bodmer \etal{bo}).
\end{description}
\begin{figure}
\epsfxsize 8.cm
\epsfbox[25 18 587 524]{fig_1.ps}
\caption[]{
\he\ profiles for $R\leq 1.2R_\odot$, in
a solar model initialized with \dhp$=3.01 \times 10^{-5}$ for various ages:
0\,Myr (dash-dot-dot-dot), 0.138\,Myr (dashed), 20\,Myr (dash-dot-dash),
49.1\,Myr (thin full), 200\,Myr (dotted) and 4550\,Myr (thick full).
}\label{fig:1}
\end{figure}
\begin{figure}
\epsfxsize 8.cm
\epsfbox[25 18 587 524]{fig_3.ps}
\caption[]{
Changes of \he\ (thick) and $^4{\rm He}/^1{\rm H}$ (thin) ratios
({\em with different scaling}) at surface with respect to time
for the solar model of \fig{fig:1}.
The dashed heavy line (49\,Myr) separates
the pre-main sequence (PMS) from the main sequence (MS).
}\label{fig:3}
\end{figure}

\section{Results and Discussion}\label{sec:rd}
For models initialized for various values of \dhp, the
\he\ profiles are plotted \fig{fig:2} for $R\gse 0.55R_\odot$
and compared to the three observed \hesw\ values.
According to these models, \dhp\ can be fitted by the quadratic polynomial:
\begin{equation}\label{eq:1}
(^2{\rm H}/^1{\rm H})_{\rm p}=2.675\times 10^{-5}+(9.364\times 10^{-2}-2.452x)x.
\end{equation}
here $x\equiv (^3{\rm He}/^4{\rm He})_\odot-4.0\times 10^{-4}$,
\he\ profiles within $0\leq R\leq 1.2R_\odot$
are plotted \fig{fig:1} at various ages of a solar model computed with
\dhp=$3.01\times 10^{-5}$. During the pre-main
sequence $^2$H is converted to $^3$He and, due to mixing,
\he\ increases through the model. At time $t=\sim 20$\,Myr the convection
zone has receded almost to its present days location, at center, the
temperature is not high enough to convert $^3$He into $^4$He via
He$^3$(He$^3$,2p)$^4$He, then \he\ is maximum there.
At $t=\sim 49$\,Myr \ie, zero age main sequence,
the flat profile of \he\ for
radius $R\loa 0.1R_\odot$, is due to the mixing in the convective
core resulting form the conversion of $^{12}$C into $^{14}$N;
then the nuclear reactions
reach their equilibrium. Around the center the increase of $^4$He
depresses \he\ and its maximum progressively reaches its present days location
around $0.3R_\odot$; the gravitational settling being more efficient for
$^4$He than for $^3$He, at surface \he\ {\em slowly increases}
until present days. As seen in \fig{fig:3}, along the evolution,
\he\ varies from $1.10\times10^{-4}$ to $4.361\times10^{-4}$ and \heh\ from
$0.0966$ to $0.0830$.

Using \eq{eq:1} the three observed \hesw\ values allow to infer:
\begin{description}
\item{(1)} \dhp$=(2.91\pm 0.19)\times10^{-5}$,
\item{(2)} \dhp$=(3.50\pm 0.43)\times10^{-5}$,
\item{(3)} \dhp$=(3.05\pm 0.36)\times10^{-5}$.
\end{description}
Assuming no systematic errors, the weighted mean of these determinations
results in:
\[
(^2{\rm H}/^1{\rm H})_{\rm p}=(3.01\pm 0.17) \times 10^{-5}.
\]
Niemann \Etal{ni} have derived a similar value but using our \hep$=0.0966$
value, they would have found \dhp$=(2.9\pm 1)\times10^{-5}$.

Our new estimate of \dhp\ overlaps within the domain of uncertainty the
Galileo result. At this point, we can thus consider that it is consistent
with Galileo. However, it rules out Jovian values
higher than \dhj$\ga 3.18\,10^{-5}$ if we assume that Jupiter is
representative of the isotopic composition of the nebula.
As mentioned in \sec{sec:in}, \dhj\ results
from a mixing of hydrogen originating from the nebula -- and thus in
protosolar abundance -- with ices more or less enriched in deuterium.
Assuming that the two reservoirs equilibrated at high temperature at the
time of the formation of the planet, the present hydrogen in Jupiter
may have been somewhat enhanced in deuterium if the amount of ices was
large enough.
This question deserves to be reexamined in the light of the most recent
models of interiors of Jupiter.
Following Hubbard \& MacFarlane (\cite{hmf}), the deuterium enhancement
is a function of the mass $M_{\rm i}$ of the ices embedded in Jupiter and of
their deuterium enrichment $f$ with respect to the protosolar value.
According to Guillot \Etal{gug},
the amount of ices should not exceed 32 Earth masses in the extreme case.
The value of the enrichment $f$ depends on the
origin of Jovian protoices. They may have been formed in the solar nebula,
which implies that
$f$ not to exceed 2.5. In such a case,
the deuterium enhancement in Jupiter is negligible.
Alternatively, ices may have originated directly from the protosolar cloud
following the scenario discussed by Lunine \Etal{lu} and would have then
kept their interstellar isotopic signature. The recent determinations of
\dh\ in water in P/Halley (Eberhardt
\etal{eb}) and in Hyakutake comet (Gautier \etal{ga})
suggest $f$ of the order of 10. This
{\it  maximum} scenario then results in a deuterium enhancement of 20\%.
Accordingly, \dhj\ should then be equal to \dhj$=(3.65\pm 0.2)\times
10^{-5}$,
entirely within the error bars of the Galileo value.

Ground based remote sensing determinations
of \dhj\ conclude to values less than
$3.0\times 10^{-5}$ (Lecluse \etal{le}), in conflict with the
{\it  maximum} enrichment scenario. The preliminary analysis of observations
from ISO suggest similar results (Encrenaz \etal{en}).
In contrast,
the reanalysis of the Voyager infrared observations by Carlson \Etal{ca}
results in \dhj\ between $3.7\times 10^{-5}$ and
$6.0\times 10^{-5}$ (Lecluse \etal{le}), a value
compatible with the {\it  maximum} enrichment scenario.
Preliminary results of HST observations of the Lyman $\alpha$
emission of Jupiter
(Ben\,Jaffel \etal{bjf}) give \dhj=$(5.9\pm 1.4)\times 10^{-5}$,
a value much higher than the upper limit of the {\it maximum} enrichment case.

In summary, we cannot yet decide whether the abundance of deuterium in Jupiter
is representative of the protosolar value or if it has been somewhat enhanced
during the planetary formation.

The lower limit of our revised pro\-to\-so\-lar \dhp\ va\-lue,
na\-me\-ly \dhp$=2.83\times 10^{-5}$,
is higher than the upper limit of the ISM determination of Linsky \Etal{li},
by a factor 1.6. This is consistent with the evolution in
4.55\,Gyr of
the deuterium abundance in the solar neighborhood, estimated from models
of chemical evolution of the Galaxy with infall of primordial composition
(Prantzos \cite{pr}).
Obtaining higher depletion factors requires different models which
invoke galactic winds in the Galaxy
(Vangioni-Flam \& Casse \cite{vc}).

\section{Conclusion}\label{sec:c}
 We propose for the protosolar deuterium abundance:
$
(^2{\rm H}/^1{\rm H})_{\rm p}=(3.01\pm 0.17)\times 10^{-5}.
$
\noindent This result is based on observed values of
\hesw\ in the solar wind and
of \dhj\ on Jupiter; it takes into account the microscopic diffusion
of elements in solar modeling. Systematic errors could occurs:
i) although not yet detected, a fractionation between $^3$He and $^4$He
could occur between the surface and the solar wind, resulting in
a small decrease of the inferred \dhp, ii) microscopic
diffusion coefficients could be revised, affecting the calculation of \hes.
The new protosolar value is consistent with the lower range of the
preliminary determination of \dhj\ in Jupiter obtained by
the Galileo mass spectrometer. The difference between our value
and the present ISM value
of Linsky \Etal{li} is compatible with some models of the chemical evolution
of the Galaxy.
Our result does not significantly changes
the sum  \dhp+\hethp$=4.07\times 10^{-5}$,
from its previous estimate of $4.1\times 10^{-5}$ by G93.
In fact, using our value \hehp$=0.0966$, this author would have found
$4.05\times 10^{-5}$.
 
\begin{acknowledgements}
We thank R. Bodmer for helpful discussions,
and the Pr. J. Geiss who referred this Letter for constructive comments.
This work
was partly supported by the GDR G131 "Structure Interne" of CNRS (France).

\end{acknowledgements}

\end{document}